\documentclass[useAMS,usenatbib]{mn2e}
\usepackage{epsfig}

\def\mnras{MNRAS}  
\def\apj{ApJ}      
\def\apjl{ApJL}    
\def\apjs{ApJS}    
\def\aap{A\&A}     
\def\nat{Nature}   

\def\aj{AJ}        

\title[Baryon Fractions: Clusters and {\it WMAP}]{Revisiting the 
Baryon 
Fractions of Galaxy Clusters:\\ A Comparison with {\it WMAP} 3-year 
Results}

\author[I. G. McCarthy et al.]{I. G. McCarthy$^{1}$\thanks{E-mail:
i.g.mccarthy@durham.ac.uk (IGM)}, R. G. Bower$^1$, and M. L. Balogh$^2$
\\
\\
$^{1}$Department of Physics, University of Durham, South Road,
Durham, DH1 3LE\\
$^{2}$Department of Physics and Astronomy, University of Waterloo,
Waterloo, ON, N2L 3G1, Canada}

\begin{document}

\date{Accepted XXXX. Received XXXX; in original form XXXX}

\pagerange{\pageref{firstpage}--\pageref{lastpage}} \pubyear{2006}

\maketitle

\label{firstpage}

\begin{abstract}
The universal baryonic mass fraction ($\Omega_b/\Omega_m$) can be 
sensitively 
constrained using X-ray observations of galaxy clusters. In this 
paper, we compare the baryonic mass fraction inferred from 
measurements of the cosmic microwave background with the gas mass 
fractions ($f_{\rm gas}$) of a large sample of clusters taken 
from the recent literature.  In systems cooler than 4 keV, $f_{\rm 
gas}$ declines as the system temperature decreases.  However, in 
higher temperature systems, $f_{\rm gas}(r_{500})$ converges to 
$\approx \left(0.12\pm0.02\right)\left(h/0.72\right)^{-1.5}$, where 
the uncertainty reflects the systematic variations between clusters 
at $r_{500}$.  This is significantly lower than the 
maximum-likelihood value of the baryon fraction from the recently 
released {\it WMAP} 3-year results.  We investigate possible 
reasons for this discrepancy, including the effects of radiative 
cooling and non-gravitational heating, and conclude that the most 
likely solution is that $\Omega_m$ is higher than the best-fit 
{\it WMAP} value (we find $\Omega_m = 0.36^{+0.11}_{-0.08}$), 
but consistent at the $2\sigma$ level. Degeneracies within the {\it 
WMAP} data require that $\sigma_8$ must also be greater than the 
maximum likelihood value for consistency between the data sets.
\end{abstract}


\section{Introduction}

For over a decade now cluster gas mass fractions as inferred from 
X-ray observations have been used as a probe of the universal 
ratio of baryon to total matter densities, $\Omega_b/\Omega_m$ 
(e.g., White et al.\ 1993; David et al. 1995; Evrard 1997; Mohr 
et al.\ 1999; Roussel et al.\ 2000; Allen et al.\ 2002; Lin et 
al.\ 2003; Ettori 2003; Allen et al.\ 2004).  Supplementing 
these gas mass fractions with constraints on $\Omega_b$ from, 
e.g., cosmic microwave background (CMB) measurements or a 
combination of Big Bang Nucleosynthesis (BBN) predictions and 
D/H measurements from high redshift quasars, therefore allows 
one to measure the total matter 
density $\Omega_m$.  The reliability of this test rests on the 
assumption that clusters have been able retain the original baryon 
inventory assigned to them in the early universe.  So-called 
``non-radiative'' cosmological simulations, which include a 
hydrodynamic treatment of the baryons but neglect sources or 
sinks such as radiative cooling, star formation, and feedback, indeed 
indicate that clusters retain nearly all their baryons until the 
present day (e.g., Frenk et al.\ 1999; Kay et al.\ 2004; Crain et 
al.\ 2006).  The same is generally true for simulations with cooling
and feedback.  Although the fraction of baryons in the hot phase 
depends strongly on the model, most recent simulations predict a 
mild increase in the hot gas fraction with cluster mass, and 
little evolution with redshift (e.g., Tornatore et al.\ 2003; 
Kravtsov et al.\ 2005; Ettori et al.\ 2006).

Although the cluster baryon fraction test has been examined 
previously in many studies, there are several good reasons for 
revisiting it.  First, new high-quality data obtained from the 
{\it Chandra} and {\it XMM-Newton} telescopes now allow us to probe 
both the surface brightness and temperature profiles of clusters 
out to large radii.  As a result, both the statistical and 
systematic observational uncertainties on the gas mass fraction 
are substantially improved.  Second, much improved (e.g., K-band) 
measurements of the stellar content of clusters are now available.  
Third, cosmological simulations can now robustly predict the 
baryon fractions within $r_{500}$, which is roughly the same 
radius the latest X-ray measurements reliably extend out to.  
Fourth, analyses of mock observations of realistic 
cosmological simulations allow us to, e.g., quantify the 
observational bias introduced by assuming strict hydrostatic 
equilibrium (HSE) in the derivation of X-ray 
gas mass fractions.  Finally, analysis of the 
recently released {\it WMAP} 3-year cosmic microwave background 
(CMB) data has yielded very tight 
constraints on the universal baryon fraction (Spergel et al.\ 
2007).  As a result, there is now an excellent opportunity to 
check for consistency (or lack of) between these two very 
different tests.  We note that Ettori (2003) previously found 
evidence for a slight discrepancy between the universal baryon 
fraction inferred from the first year {\it WMAP} data and cluster 
baryon fractions inferred from {\it BeppoSAX} X-ray data.  It is 
interesting to see whether or not this discrepancy remains when 
one makes use of the better observational data and improved 
theoretical predictions that are now available.

In this paper we analyse the best literature X-ray data in a 
homogeneous way, and compare the results with the universal baryon 
fraction inferred from the recently released {\it WMAP} 3-year 
data (Spergel et al.\ 2007).  

Unless otherwise stated, we assume a $\Lambda$CDM cosmology with 
$h=0.72$.

\section{Cluster gas fractions}

We select high quality data from a few recent studies in the 
literature where the mass profiles have been reliably estimated 
out to $\sim r_{500}$.  Vikhlinin et al.\ (2006), hereafter V06, 
have measured the gas and total mass profiles for a sample 13 
relaxed ``cool core'' observed with {\it Chandra}.  From this 
sample, we select all but two clusters\footnote{We exclude 
A2390, which V06 have demonstrated to be highly asymmetric owing 
to the presence of a set of large X-ray cavities nearly 400 kpc 
in diameter, and the low-temperature system USGC S152, for which 
V06 do not provide enough information to reconstruct its mass 
profile.}.  In addition to the above, we also select a sample of 
10 relaxed ``cool core'' clusters observed with {\it XMM-Newton} 
for which Pointecouteau et al.\ (2005), Arnaud et al.\ (2005) 
and Pratt et al.\ (2006a) (hereafter collectively referred to as 
PAP) have computed gas and total mass profiles.  Therefore, in 
total we have compiled a sample of 21 sets of mass profiles from 
19 different clusters (i.e., the samples have two clusters in 
common).

Both V06 and PAP computed the gas and total mass profiles of 
their clusters in a similar manner, by fitting parametric forms 
of the gas density and (3D) temperature profiles of the ICM 
(see eqns.\ 3 and 6 of V06 and Appendix A of Pratt et al.\ 2006a) to
the observed, projected surface brightness and temperature profiles.
The purpose of these parametric models is to produce a smooth
description of the data and hence reduce the noise in the spatial
derivatives; the exact form of the models is irrelevant as long 
as they fit the data well.  Note that the assumption of smooth 
gas in HSE can lead to a small overestimate of the baryon 
fraction (Mathiesen et al.\ 1999; Mohr et al.\ 1999; Rasia et 
al.\ 2006), which strengthens our conclusions.

We use the parametric models and associated best-fit parameters 
listed in V06 and PAP to reconstruct the observed mass profiles.  
In both 
studies the total mass distributions were fitted with the 
Navarro, Frenk \& White (1997, NFW) profile derived from cosmological
dark matter simulations.  Both V06 and PAP demonstrate that the 
NFW profile fits their data exceptionally well, with an inferred 
mass-concentration parameter relation that is in good agreement 
with the results of cosmological simulations. 

\begin{figure}
\centering
\includegraphics[width=8.4cm]{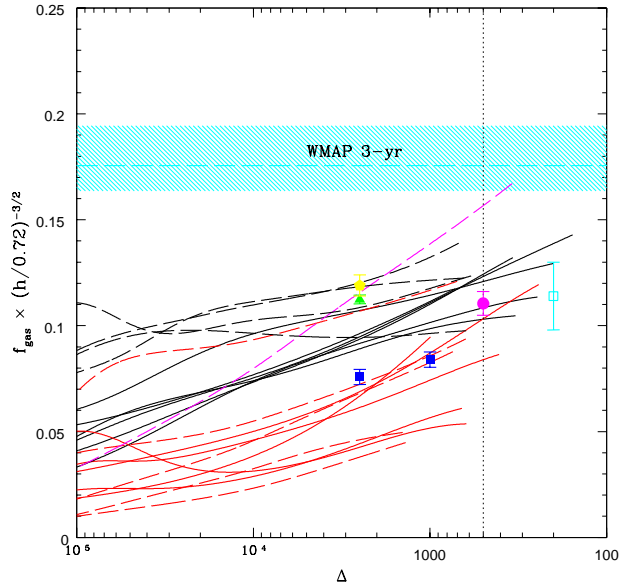}
\caption{Comparison of the observed integrated gas mass fractions
[i.e., $f_{\rm gas} \equiv M_{\rm gas}(<\Delta)/M_{\rm
tot}(<\Delta)$] as a function of overdensity ($\Delta$) with
{\it WMAP} 3-year constraints on the universal baryon fraction
$\Omega_b/\Omega_m$ (assuming a flat power-law $\Lambda$CDM
cosmology).
Solid and dashed lines represent fits to the {\it Chandra}
data of V06 and {\it XMM-Newton} data of PAP, respectively.
Red and black lines represent systems with mean spectral
temperatures below and above 4.5 keV, respectively.  The magenta
line represents PAP's fit to A1413.  Also shown (data points), 
are weighted averages of the gas mass fraction determinations of 
Ettori et al.\ (2002) (blue squares), Allen et al.\ (2004) (green 
triangle), Sadat et al.\ (2005) (magenta circle), 
LaRoque et al.\ (2006) (yellow pentagon), and Afshordi et al.\ 
(2007) (open cyan circle).  
Typically, the associated {\it statistical} measurement 
uncertainty for individual clusters is of order 10\% for 
overdensities within which the temperature and surface brightness 
profiles can be reliably measured.  The vertical dotted line 
indicates an overdensity of 500 (i.e., corresponds to $r_{500}$).  
The shaded cyan region corresponds to the 68\% confidence region 
for $\Omega_b/\Omega_m$ from {\it WMAP}, while the long dashed 
line shows the best fit value (Spergel et al.\ 2007).
 }
\end{figure}

Presented in Figure 1 is a comparison of the observed integrated 
gas mass fractions as a function of overdensity, $\Delta$, where 
$\Delta \equiv 3 M_{\rm tot}(<r) / \left[4 \pi r^3 \rho_{\rm 
crit}(z)\right]$.  In general, the clusters all show a mildly rising 
gas fraction with decreasing overdensity.  However, there is 
considerable scatter in the gas fraction at fixed overdensity that 
is worth exploring.  First, it is evident that the gas mass 
fractions measured by PAP are systematically larger than those 
measured by V06.  Given that the gas density profiles from the two 
samples are quite similar (see McCarthy et al.\ 2007), the 
implication is that the temperature profiles measured by PAP 
and V06 differ systematically from each other.  Indeed, V06 
generally find temperature profiles that decline relatively 
rapidly with radius, dropping by roughly a factor of 2 from the 
peak (at $r \sim 0.1-0.2 r_{200}$) to $0.5 r_{200}$ (see also 
Vikhlinin et al.\ 2005), while PAP find a much more gradual 
decline, with some clusters showing approximate isothermality.  
Through the equation of HSE, a flatter temperature gradient 
translates into a reduced normalisation of the total mass profile 
and hence an increased gas mass fraction.  While it 
would be useful to sort out the exact nature of the temperature 
discrepancy\footnote{It now appears that the discrepancy between 
temperature profiles derived with {\it Chandra} and those with {\it 
XMM-Newton} may be nearly resolved.  Using a larger sample of clusters 
observed with {\it XMM-Newton} and an improved model for background 
subtraction, Pratt et al.\ (2006b) now find steeper temperature 
declines at large radii which are quite similar to those of V06 
(G.~W. Pratt, private communication).  See also Vikhlinin et 
al.\ (2005).}, we point out that the typical level of difference 
between the two is relatively small (compared to the offset of 
both from the {\it WMAP} result, see \S 3).  The one exception 
to this appears to be A1413, which was observed by both V06 and 
PAP.  Within $r_{500}$, V06 find a total mass that is roughly 
50\% larger than that found by PAP.  Increasing the total mass 
of A1413 by this factor would bring it more in line with the 
other systems studied by PAP.  However, we note that good 
agreement between PAP and V06 is found for A1991, the only other 
system in common between the two samples.

\begin{figure}
\centering
\includegraphics[width=8.4cm]{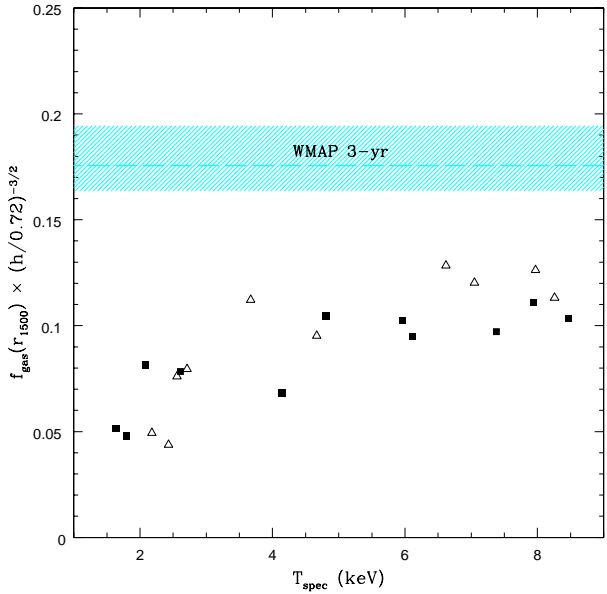}
\caption{Comparison of the observed integrated gas mass fractions
within an overdensity of $\Delta = 1500$ as a function of mean
ICM temperature with {\it WMAP} 3-year constraints on the universal
baryon fraction $\Omega_b/\Omega_m$.  Solid squares represent the
{\it Chandra} data of V06 and open triangles represent the {\it
XMM-Newton} data of PAP.
 }
\end{figure}

Fig.\ 1 also shows that the gas fraction depends on system 
temperature, which presumably reflects total system mass.  In 
particular, systems with mean temperatures below 4.5 keV (red 
lines) have systematically lower values for $f_{\rm gas}$ within 
virtually all 
overdensities compared with hotter systems (black lines).  This 
trend has been noticed previously (e.g., Mohr et al.\ 1999; Roussel 
et al.\ 2000; Neumann \& Arnaud 2001) and would seem to indicate 
that non-gravitational heat sources are relatively more important 
in mediating the properties of the ICM in low mass than in high 
mass clusters. In Figure 2, we plot the gas mass fraction within an 
overdensity of 1500 (typically corresponding to a physical radius 
of $\approx 800$ kpc) versus mean system 
temperature.  We have chosen this radius so that all the clusters 
in the sample (including the coolest systems) have reliable data.  
Note that if non-gravitational heating is unimportant at 
$r_{1500}$ it will be even less significant at $r_{500}$.  A trend 
with temperature is clearly visible at low temperatures. However, 
above $\approx 4$ keV both the {\it Chandra} and {\it XMM-Newton} 
data show no evidence for a further increase, and the gas fraction 
appears to have converged, suggesting that non-gravitational 
physics is largely unimportant for the most massive systems (at 
least out at large radii).  
This is fully consistent with previous results based on {\it ROSAT} 
and {\it ASCA} data (Roussel et al. 2000; Lin et al.\ 2003), as well 
as newer {\it XMM-Newton} data (Sadat et al. 2005).  It is also 
consistent with the findings of McCarthy et al.\ (2007), that above 
$\sim3$ keV the gas density and entropy profiles of ``cool core'' 
systems follow the scaling expected for self-similar gravitational 
heating at large radius.

We have elected to focus on the data of V06 and PAP since their 
measurements extend out to $\sim r_{500}$.  As noted earlier, 
cosmological simulations can now reliably predict the baryon 
fraction within this radius.  However, measuring the properties 
of the ICM at such large radii is not easy.  The rapidly 
declining surface brightnesses of the clusters at large radii 
means that an extremely careful treatment of the background is 
crucial.  For this reason, a number of groups have limited their 
analyses to smaller radii (higher surface brightnesses) in an 
attempt to mitigate any potential systematic effects associated 
with the background modelling.  While we are encouraged by the 
similarity of the results of V06 and PAP at large radii (note 
also that {\it Chandra} and {\it XMM-Newton} have different 
background characteristics), it is useful to check these results 
against other (mainly small-radii) data from the recent 
literature.

With the above in mind, we include some other recent X-ray gas 
mass fraction determinations (within various fixed overdensities) 
of hot clusters from the literature (data points) in Figure 1.  
In particular: the (yellow) pentagon represents a weighted 
average 
of $f_{\rm gas}(r_{2500})$ for the 38 clusters (spanning $0.14 < 
z < 0.89$) observed by LaRoque et al.\ (2006) with {\it Chandra} 
[note that LaRoque et al. actually use both {\it Chandra} X-ray 
data and {\it BIMA/OVRO} Sunyaev-Zeldovich effect data to 
constrain $f_{\rm gas}(r_{2500})$]; the (green) triangle 
represents a weighted average of $f_{\rm gas}(r_{2500})$ for the
26 clusters (spanning $0.07 < z < 0.9$) observed by Allen et al.\
(2004) with {\it Chandra}; the (blue) squares represent weighted
averages of $f_{\rm gas}(r_{2500})$ and $f_{\rm gas}(r_{1000})$
for the 22 clusters (spanning $0.01 < z < 0.1$) observed by 
Ettori et al.\ (2002) with {\it BeppoSAX}, and; the (magenta) 
circle represents a weighted average of $f_{\rm gas}(r_{500})$ 
for the 8 clusters (at $z \approx 0.5$) observed by Sadat et al.\ 
(2005) with {\it XMM-Newton}.

Although there is some scatter in the averaged $f_{\rm gas}$ 
values from these different studies, Fig.\ 1 demonstrates that 
they are, by and large, consistent with the profiles of V06 and 
PAP.  The one exception is the data of Ettori et al.\ (2002), 
which are systematically lower than the others.  Since this is 
the data Ettori (2003) used to show there was only a very small 
discrepancy between cluster baryon fractions and the first-year
{\it WMAP} results, we discuss the studies of Ettori et al.\ 
(2002) and Ettori (2003) in a bit more detail in \S 3.

X-ray observations are not the only way one can constrain the gas
mass fractions of clusters.  Increasingly, the Sunyaev-Zeldovich
(SZ) effect is also being used for this purpose.  For example,
LaRoque et al.\ (2006) found very similar results for $f_{\rm
gas}(r_{2500})$ when fitting to both {\it Chandra} X-ray data
and {\it BIMA/OVRO} SZ effect data simultaneously (yellow 
pentagon in Fig.\ 1) or when fitting to the SZ effect data alone.  
More recently, Afshordi et al.\ (2007) have stacked the SZ effect
signals of 193 clusters in the {\it WMAP} 3-year data to 
constrain the thermal energy of the ICM.  With the aid of 
hydrodynamic simulations, they converted this thermal energy into 
an estimate of the gas fraction of clusters within $r_{200}$.  
Their result, represented by the open (cyan) square in 
Fig.\ 1, is in excellent accordance with the X-ray-derived 
results.  This is very encouraging since the systematics 
involved in the derivation of this result are obviously very 
different from those associated with X-ray measurements.

\section{Implications}

In cosmological simulations the baryon fraction within $r_{500}$ has 
converged to $90-95$\% of the universal value in the case of 
non-radiative simulations (e.g., Frenk et al.\ 1999; Kay et al.\ 
2004; Crain et al.\ 2006) and slightly higher than this when 
radiative cooling and star formation is included (e.g., Kravtsov et 
al.\ 2005; Ettori et al.\ 2006).  It is only possible to drive a 
substantial fraction of the baryons beyond  $r_{500}$ if the 
energy input from non-gravitational heating (such as AGN powered 
jets and bubbles; e.g., Churazov et al.\ 2002) is comparable to the 
binding energy of the cluster gas (we quantify the energy 
requirement in \S 3.2). In the absence of such high levels 
non-gravitational heating, the baryon fractions derived at $r_{500}$ 
should be nearly representative of the universal value.

Independent measurements of the universal baryon fraction
$\Omega_b/\Omega_m$ can be derived from the power spectrum of CMB
anisotropies.   In Figs.\ 1 and 2 we show the value of 
$\Omega_b/\Omega_m$ constrained by the {\it WMAP} 3-year data (Spergel
et al.\ 2007).  From Fig.\ 1 it is immediately apparent that, with 
the exception of PAP's estimate of $f_{\rm gas}$ for A1413, the 
observational gas mass estimates do not achieve consistency with 
the {\it WMAP} measurements of the baryon fraction within any 
(observable) overdensity.  This is true despite the fact that 
the {\it XMM-Newton} data of PAP extend out to nearly $r_{500}$ 
(note 
that, typically, $r500 \sim 1$ Mpc) and slightly beyond for the 
{\it Chandra} data of V06.  At the lowest observable overdensity 
(largest radii), the data indicate a gas mass fraction that is 
roughly 40\% lower than $\Omega_b/\Omega_m$ inferred from {\it 
WMAP} 3-year data.  Fig.\ 2 demonstrates that the observed gas mass 
fractions at $> r_{1500}$ have converged for systems above $\approx 
4$ keV, and still lie well below the {\it WMAP} 3-year 
constraints.  We now examine the possible origins of this sizeable 
discrepancy.  For ease of discussion, we break up the possible 
solutions into three broad categories.

\subsection{Stars and cool baryons}  

X-ray data by itself constrains only the fraction of a cluster's mass 
in the form of hot ($T > 10^{6.5}$ K) gas.  A proper comparison 
to the universal {\it WMAP} baryon fraction therefore requires that 
we take into account the fraction of cluster's baryons locked up in 
cool gas (that doesn't emit X-rays), stars and baryonic dark 
matter.  If clusters manage to significantly cool $\approx40\%$ of 
their baryons this would potentially resolve the discrepancy 
described above.  Indeed, cosmological simulations that take into 
account the effects of radiative cooling and star formation 
demonstrate that clusters can potentially cool out such large 
quantities of their baryons (e.g., Dav{\'e} et al.\ 2002; Kravtsov 
et al.\ 2005).  But such simulations are at odds with near-infrared 
(K-band) observations of clusters, which typically indicate that 
the total (resolved) stellar mass is at most 5-14\% of the 
ICM gas mass in hot ($> 4$ keV) clusters and only slightly higher 
than this in cooler systems\footnote{We note that some 
previous studies (e.g., Ettori 2003; Voevodkin \& Vikhlinin 2004) 
estimated slightly higher stellar fractions (typically 15-20\%) 
based on V- and B-band cluster observations.  However, due to 
uncertain age and metallicity effects, conversion to stellar 
mass from the V- and B-bands is less reliable than converting from 
the K-band.} (e.g., Balogh et al.\ 2001; Lin et al.\ 2003).  
Recent deep 
optical observations limit the contribution of diffuse 
intracluster light to between $\sim10-30\%$ of the total stellar 
luminosity (e.g., Gonzalez et al.\ 2005; Zibetti et al.\ 2005).  
Low mass stars and brown dwarfs (or ``rocks'') are also likely to be 
a significant, undetected component of the mass budget, but there is 
no strong evidence that they are more abundant than expected from 
standard initial mass functions used to extrapolate the observed 
stellar mass functions (Fuchs, Jahreiss \& Flynn 1998; Gizis
et al.\ 2000; Lucas et al.\ 2005; Levine et al. 2006).  Radio, 
infrared, and ultraviolet surveys for atomic, molecular, and 
ionised gas (respectively) limit the contribution of cool
($T < 10^{5.5}$ K) diffuse baryons to less than a percent or
so of the hot X-ray-emitting ICM (e.g., O'Dea et al.\ 1998; Donahue
et al.\ 2000; Edge 2001; Edge \& Frayer 2003; Bregman et al.\
2006).  Finally, Ettori (2003) have suggested that a significant 
fraction of cluster baryons could lie at temperatures between 
$10^{5} < T < 10^{7}$ K and they cite possible evidence for this 
in the form the observed ``soft X-ray excess''.  However, it now 
appears that this excess was the result of inadequate modelling of 
the Galactic foreground (Bregman \& Lloyd-Davies 2006). 
Furthermore, it is unclear physically how such large quantities of 
gas in this temperature range could resist cooling down to $10^4$ 
K or being mixed or heated to the ambient ICM temperature.

Furthermore, one can place limits on any hidden cold baryonic 
component by considering its gravitational effects on the 
properties of the ICM.  In particular, if the hidden material is 
centrally concentrated, as one would expect if its origin is 
linked to cooling, this will deepen the potential well which, in 
turn, will heat the ICM through simple gravitational compression.  
The observed temperature profiles can therefore be used to place 
constraints on the amount of hidden cold material in clusters.  We 
have tested this as follows.  In particular, we assume the dark 
matter follows a NFW distribution with a typical concentration 
parameter $c_{200} = 4$.  We further assume the gas 
density profile follows a distribution typical of observed ``cool 
core'' clusters (see V06; McCarthy et al.\ 2007).  Finally, we try 
different hidden cold mass distributions.  The predicted ICM 
temperature distributions are computed by placing the gas in HSE.  
In the case where there is no hidden component, we find the 
resulting temperature profiles quantitatively match the observed 
ones.  This is not unexpected, given the results of McCarthy et 
al.\ (2007) (see \S 3.1 of that study).  However, the predicted 
temperature profiles are quite sensitive to the addition of a 
centrally-concentrated cold component.  In particular, if we 
require that the predicted temperature profiles are within the 
observational uncertainties (typically, binned temperature 
profiles have a 10\% measurement uncertainty), we can rule out a 
cold component weighing $> \ \approx 3\times10^{12} M_{\odot}$ 
within the central 50 kpc or $> \ \approx 5\times10^{12}M_{\odot}$ 
within the central 100 kpc. (Note that our results at a 
fixed radius are not sensitive to the distribution of the cold 
component, only to the total enclosed mass within that radius.)
This is small compared to the $\sim 5\times10^{13} M_\odot$ 
required to resolve the cluster vs. CMB discrepancy for a massive 
$10^{15} M_\odot$ cluster.  The only way to reconcile the cluster 
vs. CMB discrepancy and maintain consistency with the observed 
temperature profiles is if the hidden component is very 
spatially-extended.  However, such a configuration seems 
physically contrived.

It therefore appears that the stellar/cool baryon contribution to the 
total cluster baryon budget too small to account for the cluster 
vs. CMB discrepancy.  In particular, the resolved stellar/cold 
baryon component is a approximately factor of 3 smaller than 
required.  Moreover, we have demonstrated that any hidden component 
cannot be very massive, otherwise this would violate the observed 
temperature profiles of relaxed clusters.

\subsection{Non-gravitational heating}

As discussed above, the small fraction of cooled baryons observed, 
relative to the predictions of cooling-only simulations, implies 
that some form of non-gravitational heating (``feedback'') is at 
work in the ICM.  Without significant feedback to prevent this 
overcooling, theoretical models are unable to account for the 
observed cut-off of the galaxy luminosity function at the bright 
end or the fact that BCGs are, by and large, ``red and dead'' (e.g., 
Benson et al.\ 2003; Bower et al.\ 2006), nor would it be possible 
to explain the lack of X-ray emission from intracluster gas with 
temperatures below about 1 keV (e.g., Peterson et al.\ 2003).
Non-gravitational heating also appears to be necessary to account 
for the X-ray scaling properties of clusters (e.g., Kaiser 1991; 
Evrard \& Henry 1991; Babul et al.\ 2002; Voit et al.\ 2002; 
McCarthy et al.\ 2004).  Injecting thermal energy into the ICM will 
cause it to expand (while leaving the dark matter in-tact) and 
therefore will reduce the gas mass fraction 
within a given radius.  Can this heating explain the discrepancy 
between cluster and CMB measurements?  

To test this, we have computed the bulk energy required to 
transform clusters with the universal baryon fraction (all in hot gas) 
into the observed systems.  For our {\it baseline model clusters}, 
we assume the gas traces the total matter at all 
radii and, therefore, within any radius the integrated gas mass 
fraction is always the best-fit {\it WMAP} 3-year value 
$\Omega_b/\Omega_m = 
0.176$.  The total mass profiles (which are dominated by dark 
matter) are assumed to be the same as those measured by V06 and 
PAP, thus we construct a baseline model cluster for each of the 
observed systems.  The temperature profiles are determined by 
placing the gas in HSE.  Calculating the total energy of the gas in 
these systems (i.e., the summation of the total internal and 
potential energies) is then straightforward.  For the {\it 
observed} 
systems, we extrapolate the gas and total density profiles beyond 
the maximum radius to which they can be observed, until the 
integrated gas mass fraction is the universal ratio.  (Note 
that we also extrapolate the baseline models to this radius to 
ensure that both the baseline and observed systems have the same 
integrated gas and total masses.).  We assume the total density 
profiles continue to follow the NFW form fit by V06 and PAP.  We 
try various different powerlaw extrapolations for the gas density 
profiles with plausible indices ranging from $0$ to $-2.5$.  We find 
that the minimum energy required to convert the baseline models into 
the observed systems occurs when the gas density is constant with 
radius outside the maximum observable radius.  This configuration 
is perhaps unlikely, but it does provide a useful {\it lower limit} 
to the amount of heating required.  Like the baseline models, we 
place the gas in the observed systems in HSE.  Reassuringly, we 
verify that the resulting temperature profiles are in good 
agreement with the observed profiles.  The total (minimum) energy 
required to heat the ICM is just the difference of the total 
energy of the observed and baseline model systems.

For hot ($T_{\rm spec} > 4$ keV) clusters, we find that a 
substantial amount of energy is required, ranging between 
$6-45\times10^{62}$ ergs with a mean of $\approx 2.2\times10^{63}$ 
ergs (assuming $\rho_{gas}$ is constant outside the maximum 
observable radius - i.e., the minimum required energy).  It is 
interesting to compare this minimum energy estimate with the energy 
that can potentially be deposited by AGN, the most powerful source 
of non-gravitational heating we know of in clusters of galaxies.  
(We note that the energy available from AGN far exceeds that 
available from stellar feedback.)  We estimate the amount of AGN 
energy available to be tapped as follows.  First, we convert Lin 
et al.\ 
(2003)'s observed relationship between stellar mass fraction and 
total mass within $r_{500}$ (see equation 10 of that 
study\footnote{We have slightly adjusted this relation by scaling up 
the total masses of Lin et al.\ by 1.26 to account for the 
normalisation difference between the {\it ASCA} total 
mass-temperature relation assumed by Lin et al.\ and the more 
accurate {\it Chandra} relation measured by V06.}) into a 
stellar---total mass relation [i.e., $M_{\rm 
star}(r_{500})-M_{500}$] assuming a gas mass fraction of 0.12 
within $r_{500}$ (see Fig.\ 1).  This relation can be converted into 
a relationship between the total mass in black holes within 
$r_{500}$ by (optimistically) assuming that the entire stellar mass 
is contained in bulges and adopting the black hole---bulge mass 
($M_{\rm BH}-M_{\rm bulge}$) relation of H{\"a}ring \& Rix (2004).  
This leads to the following $M_{\rm BH}(r_{500})-M_{500}$ relation:
\begin{eqnarray}
\log_{10}[M_{\rm BH}(r_{500}) \ (M_\odot)] = 1.12 \log_{10}
\biggl[\frac{f_{\rm gas}(r_{500})}{0.12} \biggr] \\
\ \ \  + \ \ 0.84 \log_{10} \biggl[\frac{M_{500}}{5\times10^{14} 
M_\odot}  \biggr] + 10.229 \nonumber
\end{eqnarray}

\begin{figure}
\centering
\includegraphics[width=8.4cm]{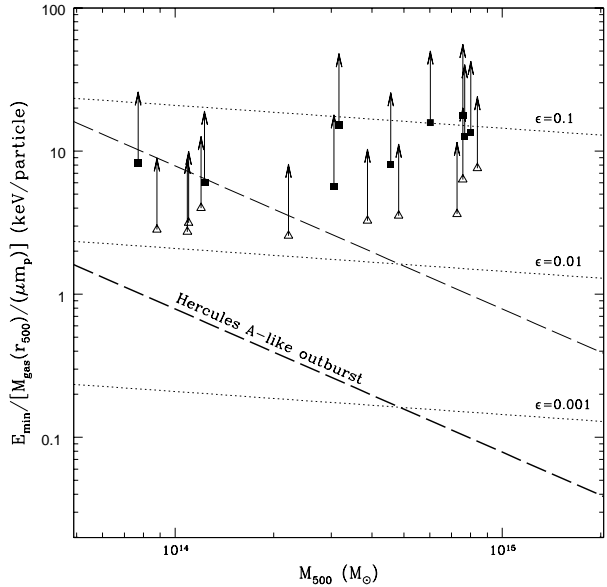}
\caption{Comparison of the minimum specific energy required to
reduce cluster gas mass fractions from the universal value to the
observed values with estimates of the amount of AGN energy
available.  Solid squares represent the {\it Chandra} data of V06
and open triangles represent the {\it XMM Newton} data of PAP.  The
three dotted lines represent theoretical estimates of $E_{\rm AGN}$,
defined as $\epsilon M_{\rm BH} c^2$ where $\epsilon$ is the
efficiency factor.  The thick dashed line shows the amount of energy
deposited into the ICM by the AGN in Hercules A, which is the most
energetic AGN outburst known (Nulsen et al.\ 2005).  The thin
dashed line corresponds to the case where a given cluster
experiences 10 such Hercules-like outbursts over its lifetime.
 }
\end{figure}

Finally, this is converted into an estimate of amount of AGN energy 
available via $E_{\rm AGN} = \epsilon M_{\rm BH}(r_{500}) c^2$.

In Figure 3, we present a comparison of the minimum {\it specific} 
energy required to resolve the cluster vs. CMB discrepancy
with the energy available to be tapped in black holes.  (To 
facilitate comparison with the observations, we calculate the 
specific energy by dividing the total required energy by the 
typical {\it observed} mass of gas within r500.  In 
particular, we assume $M_{\rm gas}(r_{500}) = 0.12 M_{500}$; see 
Fig.\ 1.).  In order to 
explain the most massive systems, we calculate that a minimum 
energy of $\sim 10$ keV per particle is required.  If one adopts 
an efficiency of $\epsilon=0.1$, which is approximately the 
efficiency predicted by standard radiatively efficient accretion 
disk models (e.g., Shakura \& Sunyaev 1973), there is potentially 
just enough energy available in black holes distributed throughout 
$r_{500}$ to account for the observed gas mass fractions.  (We use 
the term `potentially' since we remind the reader that we have 
calculated the {\it minimum} energy required and furthermore have 
made optimistic assumptions about the mass of black holes 
available to heat the ICM.)  This result agrees quite well with 
the more detailed calculations of Bode et al.\ (2007) 
(that include, e.g., the effects of asphericity and substructure), 
when one normalises their estimated required total energy to the 
same total mass of gas assumed above.

However, modelling of AGN-blown X-ray cavities suggests the typical 
cluster black hole efficiency is actually much lower than 
$0.1$.  The most energetic AGN outbursts known, in Hercules A 
(Nulsen et al.\ 2005) and MS0735.6+7421 (McNamara et al.\ 2005), 
have mean powers of $\approx 1.6-1.7 \times 10^{46}$ ergs s$^{-1}$.  
The typical age of such outbursts is $\approx$ 100 Myr, 
corresponding to a total energy of few times $10^{61}$ ergs or a 
specific energy of a few tenths of a keV per particle (see the 
thick dashed line in Fig.\ 3).  This falls nearly two orders of 
magnitude short of the required minimum to reduce a massive 
cluster's baryon fraction from the universal {\it WMAP} value to 
the observed fraction.  Therefore, even if a typical cluster 
experiences 10 such powerful outbursts over its lifetime (say, 
once per Gyr over 10 Gyr) the energy injected into the ICM 
still falls short of the minimum required energy by up to an order 
of magnitude.

We therefore conclude that AGN heating is a highly implausible, but 
perhaps not impossible, solution to the cluster vs. CMB 
discrepancy. In addition to the exceptionally large energy 
requirements, we point out that the heating must be distributed in 
just such a way as to explain the convergence trend in Fig.\ 2 and 
the fact that the ICM properties at large radii in massive 
clusters follow the {\it gravitational} self-similar scalings 
(McCarthy et al.\ 2007).

\subsection{Different cosmological parameters: $\Omega_m$ and $h$}

The observed gas fractions are proportional to $h^{-1.5}$, while the
{\it WMAP} constraint is independent of $h$.  Thus, adopting a 
lower value of
$h\approx 0.55$ would bring these two results into agreement.
Indeed, this is why similar analyses by Roussel et al. (2000) and Sadat
et al. (2005), who adopt $h=0.5$, find higher gas fractions than we 
have shown here.  However, the large body of independent evidence 
in favour of $h>0.6$ (e.g. York et al. 2005; Jones et al. 2005; 
Riess et al. 2005; Ngeow \& Kanbur 2006) makes this solution seem 
unlikely. 

We can use the observed baryon content of clusters to 
reverse-engineer the universal total matter density, $\Omega_m$, 
via:

\begin{equation}
\Omega_m = \frac{b_{\rm dep}(r_{500}) \Omega_b}{b_{\rm HSE} f_{\rm 
gas}(r_{500}) [1 + M_{star}(r_{500})/M_{\rm gas}(r_{500})]}
\end{equation}

\noindent where $b_{\rm dep}(r_{500})$ is the baryon depletion 
factor within $r_{500}$ and $b_{\rm HSE}$ is the observational bias 
introduced by assuming strict HSE (i.e., ignores pressure 
support due to turbulent motions), both of which can be 
estimated using gasdynamic cosmological simulations.  

Adopting the {\it WMAP} baryon density of $\Omega_b h^2 = 
0.0223^{+0.0007}_{-0.0009}$ (which is in good agreement with the 
latest QSO constraints; see O'Meara et al.\ 2006), $b_{\rm 
dep}(r_{500}) = 0.95 \pm 0.05$ (Kravtsov et al. 2005; Ettori et 
al.\ 2006; Crain et al.\ 2006), $b_{\rm HSE} \approx 0.9$ (Nagai 
et al.\ 2007), $M_{star}(r_{500})/M_{\rm gas}(r_{500}) = 0.10 
\pm 0.05$ (see \S 3.1), and $f_{\rm gas}(r_{500}) = 0.115 \pm 
0.015$ (see Fig.\ 1 - note that this spans both the {\it 
Chandra} and {\it XMM-Newton} results at $r_{500}$), we find:

\begin{eqnarray}
\Omega_m = 0.36^{+0.11}_{-0.08} \nonumber
\end{eqnarray}

\noindent This is larger than the best-fit {\it WMAP} 3-year value 
of $\Omega_m = 0.238 \pm 0.03$, but is in good agreement with 
several previous X-ray analyses (e.g., Allen et al.\ 2002; Lin et 
al.\ 2003; Hicks et al.\ 2006).  

However, our findings differ somewhat from those of Ettori (2003).  
He found evidence for only a marginal discrepancy between his 
cluster baryon fraction estimates (inferred from {\it BeppoSAX} 
X-ray data) and the first-year {\it WMAP} results (Spergel et al.\ 
2003).  This is somewhat surprising since Ettori (2003) used the 
data of Ettori et al.\ (2002), who actually find {\it lower} gas 
mass fractions than we have assumed above.  A likely explanation 
for this discrepancy is that: (1) he assumed a slightly higher 
stellar mass fraction; (2) he invoked a large fraction of the 
cluster baryons being hidden at ``warm'' temperatures of $10^{5} < 
T < 10^{7}$ K; and (3) he extrapolated his results well beyond the 
maximum observable radius ($\sim r_{1000}$) to the virial radius 
assuming a King total mass profile for several of his systems.  
The first two are inconsistent with recent, more reliable data 
(see \S 3.1).  Finally, extrapolation of the data out to the 
virial radius is highly uncertain.  Since simulations and 
observations now reliably extend out to $r_{500}$, such large 
extrapolations are no longer required.

In addition to the agreement with several previous X-ray studies, 
our results on $\Omega_m$ also agree with Sloan Digital Sky 
Survey (SDSS) measurements of the Lyman alpha forest power 
spectrum (Viel \& Haehnelt 2005), the baryon acoustic peak of 
luminous red galaxies (Eisenstein et al.\ 2005), and the power 
spectrum of galaxies (Tegmark et al.\ 2004).  We also note that 
combining {\it WMAP} 3-year joint constraints on $\Omega_m h^2$ 
and $\sigma_8$ (which are degenerate, see below) 
with weak lensing cosmic shear measurements (Hoekstra et al.\ 2006) 
results in an increased best-fit value for $\Omega_m$ that is in 
good agreement with our results (see Fig.\ 7 or Spergel et al.\ 
2007).  Finally, Li et al.\ (2006) have also recently reported that 
the ``low'' value of $\Omega_m$ reported by Spergel et al.\ 
(2007) is in discord with the number of observed strong lensing 
giant arcs, however a fiducial flat model with $\Omega_m = 0.3$ 
and $\sigma_8 = 0.9$ is able to match the lensing data.

On the other hand, several other cosmological tests 
tend to support the ``low'' value of $\Omega_m$ advocated by 
Spergel et al.\ (2007).  These include measurements of the 
motions within the local supercluster (Mohayaee \& Tully 2005), 
the power spectrum of galaxies in the 2dFGRS (Cole et al.\ 
2005), and the mass function of galaxy clusters in the HIFLUGCS 
X-ray flux-limited sample (Reiprich 2006).  

Clearly, the relatively large scatter in the reported best fit 
values of $\Omega_m$ warrants further detailed investigation.   
While our own estimate of $\Omega_m = 0.36^{+0.11}_{-0.08}$ is 
marginally inconsistent with the {\it WMAP} 3-year constraints 
(and some of the other tests that support the {\it WMAP} 
result), we have argued that this discrepancy is a small price to 
pay for obtaining agreement with the observed cluster gas mass 
fractions.  Finally, we point out that the discussion carries 
implications beyond the precise value of $\Omega_m$.  Since 
$\Omega_m$ and $\sigma_8$ are strongly degenerate in the CMB 
data, our results suggest that $\sigma_8$ is larger than the 
best-fit value of $0.74^{+0.05}_{-0.06}$ derived from {\it WMAP} 
data alone.  It is interesting to note that a model with 
$\Omega_m = 
0.28$ and $\sigma_8 = 0.8$ is within $\sim 1-$sigma of both the 
cluster gas mass fraction data and the {\it WMAP} joint 
constraints on these parameters.

\section{Conclusions}

Recent, good quality observations of massive clusters with {\it 
Chandra} and {\it XMM-Newton} put strong observational constraints
on the gas mass fraction in massive clusters at large radius. In 
many cases the new data allow the fraction to be constrained out to 
$r_{500}$ (the radius at which the cluster density contrast is 
500). Simulations of clusters suggest that the cluster gas fraction 
at this radius should closely reflect the average baryon mass 
fraction in the universe as a whole.

We find $f_{\rm gas}(r_{500}) \approx 
\left(0.12\pm0.02\right)\left(h/0.72\right)^{-1.5}$, 
where the uncertainty reflects the systematic variations between 
clusters. This is lower than the best fit to the {\it WMAP} 3-year 
result of $f_b=0.176\pm0.02$. 

We consider whether the discrepancy could be due to a large 
fraction of the cluster gas cooling to form stars and cold gas 
clouds, or whether it could be due to strong non-gravitational 
heating transporting $\sim30$\% of the cluster gas outside 
$r_{500}$. Observational limits on the stellar and cold gas content 
of clusters appear to rule out the first possibility.  In order to 
investigate the second, we compute the energy budget required to
rearrange the cluster gas. The energy required significantly exceeds
the plausible energy input from black holes, unless their mass
accretion history is always associated with efficient jet production.

The most likely explanation is that $\Omega_m$ lies in the range 
$0.36^{+0.11}_{-0.08}$.  Such a relatively high $\Omega_m$ lies 
slightly above the 68 per cent confidence limits from {\it WMAP}, 
but is consistent with the current data at the $2\sigma$ level. 
Since the measurements of $\Omega_m$ and $\sigma_8$ are highly 
correlated in the {\it WMAP} analysis, this means that $\sigma_8$ 
is also likely to be higher than the formal best-fit value of 
$0.74$.  This has important implications for the abundance of 
collapsed objects prior to re-ionisation (Reed et al.\ 2007).

\section*{Acknowledgments}

The authors would like to thank Carlos Frenk and David Hogg for 
helpful discussions and the anonymous referee for 
suggestions that improved the paper. IGM acknowledges support 
from a NSERC 
Postdoctoral 
Fellowship and a PPARC rolling grant for extragalactic astronomy and 
cosmology at the University of Durham. RGB acknowledges the support 
of a PPARC senior fellowship.  MLB acknowledges support from a NSERC
Discovery Grant.

\bsp

\label{lastpage}

\end{document}